\documentclass[]{article} 

\usepackage{fullpage}

\newif\ifediting
\editingtrue  

\usepackage[utf8]{inputenc}
\usepackage[T1]{fontenc}

\usepackage{multirow,arydshln,multicol, wrapfig}
\usepackage{color,graphicx}
\usepackage[hyphens]{url}

\usepackage{fullpage}

\newcommand{\furl}[1]{\footnote{\scriptsize \url{#1} \normalsize}}

\newcommand{\anonymisation}[2]{#1}

\begin{document}

\title{Un jeu à débattre pour sensibiliser à l'Intelligence Artificielle\\ dans le contexte de la pandémie de COVID-19}


\author{Carole Adam\\
Univ. Grenoble-Alpes, LIG, France\\
\url{carole.adam@imag.fr}
\and 
Cédric Lauradoux\\
INRIA Privatics, Grenoble, France\\
\url{cedric.lauradoux@inria.fr}
}

\date{Septembre 2022}

\maketitle

\begin{abstract}


L'Intelligence Artificielle est de plus en plus omniprésente 
dans nos vies. On lui délègue nombre de décisions importantes : accès aux formations supérieures, détermination de peines de prison, conduite autonome de véhicules... Nous formons des ingénieurs ou chercheurs spécialistes de ce domaine, mais la population générale a très peu de connaissances sur le sujet et est très sensible aux idées (plus ou moins fausses) véhiculées par les médias : une IA objective, infaillible, qui sauvera le monde ou au contraire le conduira à sa perte. Nous pensons donc, comme souligné par l'UNESCO, qu'il est essentiel que la population ait une compréhension basique du fonctionnement de ces algorithmes, pour choisir en connaissance de cause de les utiliser (ou pas). Pour cela nous proposons un jeu sérieux\footnote{Ce jeu a été présenté dans une session spéciale sur l'éthique et la gestion de crise, \anonymisation{à la conférence ISCRAM 2022 \cite{adam2022serious}}{\cite{anonyme}}} à destination des lycéens, sous la forme d'un débat citoyen ayant pour but de choisir une solution d'IA pour contrôler une épidémie.
\\ \textbf{Mots-clés: } Jeu sérieux, Intelligence Artificielle, éducation, COVID-19
\end{abstract}

\section{Introduction}


L'Intelligence Artificielle est de plus en plus omniprésente dans nos vies. On lui délègue nombre de décisions importantes : accès aux formations supérieures, détermination du risque de récidive et choix des peines de prison, conduite autonome de véhicules, tri automatique de CV avant recrutement... Elle aurait en effet pour avantage de prendre des décisions objectives débarrassées des biais humains, et surtout des décisions plus rapides sur la base de masses de données difficilement analysables par un humain. Cependant, même si les technologies d'IA peuvent offrir de nombreux bénéfices, elles ne sont pas du tout dépourvues de biais. Au contraire elles 
reproduisent et amplifient les biais humains présents dans les données qui les nourrissent. Elles posent ainsi de nombreux risques : potentiels biais et erreurs, discrimination, ou encore mauvaise utilisation. La crise sanitaire a été l'occasion de prendre conscience d'une partie de ces risques avec une surveillance accrue facilitée par l'IA, l'utilisation élargie de la reconnaissance faciale, ou encore de drones.

Il est donc nécessaire pour les personnes soumises aux décisions de ces algorithmes d'en comprendre le fonctionnement, pour choisir en connaissance de cause de les utiliser (ou pas). L'UNESCO~\cite{unesco} souligne d'ailleurs cette nécessité. 
Or les formations à l'IA concernent pour l'instant surtout des ingénieurs ou chercheurs spécialistes de ce domaine. La population générale, elle, a très peu de connaissances sur le sujet et est très sensible aux idées (plus ou moins fausses) véhiculées par les médias : une IA objective, infaillible, qui sauvera le monde ou au contraire le conduira à sa perte. 
Nous proposons donc un jeu sérieux \anonymisation{\cite{adam2022serious}}{\cite{anonyme}} à destination des lycéens, sous la forme d'un débat citoyen ayant pour but de choisir une solution d'IA pour contrôler une épidémie.
Les participants incarnent les rôles de citoyens avec différents profils dans une ville (pas si) futuriste. Ils doivent débattre au sujet de nouvelles technologies d'IA qui pourraient être adoptées dans leur ville pour lutter contre une épidémie. Ce jeu fait partie d'une série plus large de jeux à débattre, chacun concernant une application sociétale différente (santé, transports...) de technologies d'IA. Cette série de jeux-débats cible les adolescents de niveau collège ou lycée. Le débat est toujours suivi d'un débriefing permettant aux joueurs de partager leur expérience, d'en apprendre plus sur l'IA, et de discuter de ses implications dans leur vie réelle. 

Dans cet article, nous discutons d'abord (Section~\ref{sec:educ}) de la nécessité d'éducation du grand public au sujet de l'IA, et de l'apport des jeux sérieux dans ce domaine. La Section~\ref{sec:jad} présente notre jeu à débattre. La Section~\ref{sec:solutions} se focalise plus en détail sur les solutions d'Intelligence Artificielle que le jeu permet d'introduire. La Section~\ref{sec:debrief} insiste sur l'importance du débriefing pour obtenir un impact pédagogique. Enfin la Section~\ref{sec:rex} offre des retours d'expérience tirés des premières sessions avec notre jeu, et la Section~\ref{sec:cci} conclut l'article.

\section{Nécessité d'éducation} \label{sec:educ}

\subsection{Éducation à l'Intelligence Artificielle}

Pendant la pandémie de COVID-19, l'importance d'éduquer la population est devenu encore plus apparente. En effet, le manque d'information, voire la désinformation diffusée sur les réseaux sociaux, peut avoir des conséquences graves. Ainsi, les théories du complot et autres \emph{fake news} ou \emph{infox} se propagent rapidement et peuvent augmenter la réticence envers, voire le rejet des vaccins \cite{nieves2021infodemic,lu2022covid}, avec des conséquences néfastes sur la santé. En outre, même si l'Intelligence Artificielle peut aider à lutter contre le COVID-19, il existe également des dangers à éviter : \cite{naude2020} insiste sur le besoin d'un équilibre prudent entre la confidentialité des données et la santé publique. 
Tout le monde doit être conscient des avantages et des dangers potentiels de l'Intelligence Artificielle, que ce soit dans le contexte spécifique de la pandémie ou plus généralement.

Cette exigence d'éducation à l'IA est particulièrement mise en avant dans le dernier rapport de l'UNESCO \cite[notre traduction]{unesco}. Cette éducation ne devrait pas concerner uniquement les technologies sous-jacentes, mais également la valeur et l'utilisation des données personnelles, ainsi que les implications éthiques des technologies d'IA. 
\footnotesize
\begin{quote}
    ''parce que vivre dans des sociétés numériques exige de nouvelles pratiques éducatives, une réflexion éthique, une pensée critique, des pratiques de conception responsables et de nouvelles compétences, au vu des conséquences sur le marché de l’emploi, l’employabilité et la participation citoyenne'' [p.18]

    
    ''La sensibilisation et la compréhension par le public des technologies d’IA et de la valeur des données devraient être favorisées par une éducation ouverte et accessible, l’engagement citoyen, l’éducation aux compétences numériques et à l’éthique de l’IA, [...]
    de sorte que tous les membres de la société puissent prendre des décisions éclairées concernant leur utilisation des systèmes d’IA, et soient protégés contre toute influence indue.'' [p.24]
    
    
    ''Les États membres doivent promouvoir des programmes généraux de sensibilisation aux avancées de l’IA, notamment concernant les données et les possibilités et défis découlant des technologies d’IA, les répercussions des systèmes d’IA sur les droits de l’homme, [...] et leurs implications. Ces programmes devraient être accessibles aux spécialistes comme aux non-spécialistes. Les États membres doivent développer des programmes d'éthique de l'IA pour tous les niveaux, et promouvoir la collaboration entre l'éducation aux compétences techniques de l'IA et l'éducation aux aspects humanistes, éthiques et sociaux de l'IA.'' [p.33]
    
\end{quote}
\normalsize

\subsection{Jeux sérieux}

Dans le but d'impliquer l'ensemble de la population dans l'apprentissage de l'IA et la compréhension de son utilisation pendant la crise du COVID-19, nous pensons que les jeux sérieux \cite{crookall2010serious} peuvent être un outil puissant. Les jeux sérieux ont été utilisés avec succès dans différents domaines : éduquer la population aux comportements à adopter en cas de catastrophes naturelles, par exemple les inondations \cite{rebolledo2009societal, taillandier2018games} ; sensibiliser au changement climatique et aux actions possibles \cite{wu2015climate, switch2022} ; former les compétences médicales et chirurgicales \cite{graafland2012systematic} ou les compétences communicatives des futurs médecins \cite{jackson2011teaching} ; ou plus récemment expliquer la pandémie et les mesures sanitaires pour améliorer leur acceptabilité \cite{cottineau2020, adam2022finding}.


Les jeux sérieux n'ont pas besoin d'être informatisés, et nous utilisons ce terme pour désigner tout jeu d'apprentissage, qui utilise des éléments ludiques pour atteindre un objectif pédagogique. Il existe différentes formes de jeux sérieux, des simulations informatiques \cite{taillandier2018games}, aux jeux de société ou de cartes \cite{wu2015climate}, aux récits interactifs, aux jeux de rôles \cite{jackson2011teaching}, ou aux débats \cite{simonneaux2001role}. Compte tenu de notre objectif déclaré d'impliquer le grand public, nous avons conçu un jeu ''débranché'' auquel tout le monde peut jouer, quelle que soit son éducation, sans ordinateur. Par ailleurs, nous souhaitons développer les capacités d'argumentation des futurs citoyens, pour leur permettre de faire des choix éclairés. Nous proposons donc un jeu sérieux sous la forme d'un jeu de rôle et de débat. La conception de notre jeu est expliquée dans la section suivante.

\section{Notre jeu sérieux : l'IA contre une épidémie à Wafer city} \label{sec:jad}

\subsection{\anonymisation{L'Arbre des Connaissances}{Association de médiation scientifique}}

Notre jeu fait partie d'une série plus large de jeux-débats sur l'IA dans divers domaines, développés par une association de médiation scientifique \anonymisation{l'Arbre des Connaissances (ADC)\footnote{Site de l'Arbre des Connaissances : \url{https://arbre-des-connaissances-apsr.org/}}}{(ANONYMISEE)}, dédiée à l'ouverture de la science vers les citoyens. Le but de cette série de jeux-débats est d'initier les adolescents au débat citoyen sur des questions d'actualité mêlant sciences et société.
Ces jeux se déroulent tous dans la même ville virtuelle (plus ou moins futuriste), appelée Wafer city, une ville de taille moyenne avec un million d'habitants. La moitié de la population vit en centre-ville, l'autre moitié en zone péri-urbaine ou rurale. Dans chaque jeu de la série, une question différente est posée au conseil municipal, dans lequel sont représentés 5 rôles joués par les participants au jeu. Les deux premiers épisodes de la série étaient consacrés à l'IA dans les transports (octobre 2018) et à l'IA dans la santé (février 2019). 
L'épisode décrit ici a été développé par les deux auteurs avec l'aide de deux membres de l'association. La phase de conception s'est étalée sur environ 1 an à partir de février 2021, avec de premières expérimentations en octobre 2021, et la sortie officielle en mars 2022.




\subsection{Séquence pédagogique d'un jeu-débat}
%

\begin{wrapfigure}[1]{L}{0.2\textwidth}
    \centering
    \includegraphics[scale=0.07]{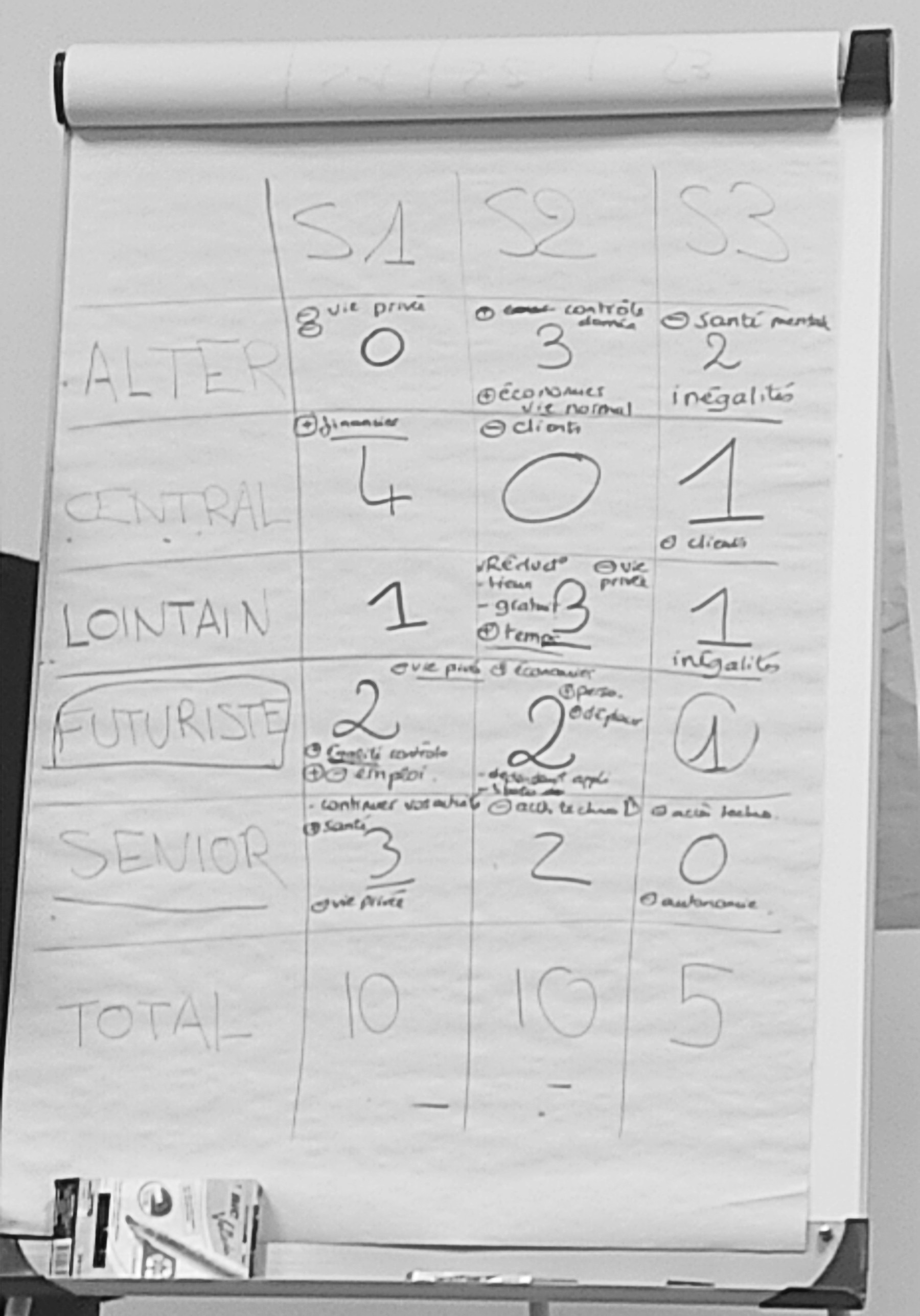}
    \caption{Tableau} 
    \label{fig:board}
\end{wrapfigure}

La séquence pédagogique pour une session de jeu-débat est la suivante~:
\begin{enumerate}
    \item La session commence par une introduction des modérateurs, qui se présentent, donnent quelques définitions sur l'IA, et exposent le scénario du jeu (cf paragraphe~\ref{ssec:scenario}). 
    \item Les participants sont répartis en cinq groupes et prennent connaissance du rôle qu'ils vont incarner au conseil municipal (cf paragraphe~\ref{ssec:roles}).
    \item Trois solutions sont présentées l'une après l'autre (cf Section~\ref{sec:solutions}). Pour chaque solution, les participants discutent avec leur groupe de ses avantages et inconvénients, selon le point de vue du rôle qu'ils représentent. 
    \item Chaque groupe distribue alors un total de 5 points entre les solutions, et fournit oralement ses scores et ses arguments, que les modérateurs reportent sur le tableau de jeu (voir un exemple Figure~\ref{fig:board}).
    \item La solution gagnante est celle qui cumule le plus de points en additionnant ceux attribués par les 5 groupes. Les résultats sont annoncés, discutés et débattus. Certains groupes peuvent être amenés à changer leurs scores en conséquence, ce qui peut modifier le résultat (voir un exemple de débat en Section~\ref{ssec:exdeb}).
    \item Après le débat, les modérateurs conduisent une phase de débriefing, qui est essentielle à l'apprentissage \cite{crookall2010serious}. Le contenu de cette phase est présenté en détails en Section~\ref{sec:debrief}.
    \item L'enseignant de la classe récupère de la documentation et des références pour approfondir le sujet avec ses élèves en classe ultérieurement (voir la liste du matériel de jeu au Paragraphe~\ref{ssec:matos})
\end{enumerate}

\subsection{Scénario du jeu : épidémie à Wafer city} \label{ssec:scenario}

Dans notre jeu, Wafer city est menacée par une épidémie. Chaque citoyen de la ville dispose d'un dossier médical numérique (contenant les résultats d'analyses, les traitements en cours...) géré par la ville et à la disposition de ses décideurs politiques, qui doivent prendre les meilleures mesures possibles afin de limiter la propagation de l'épidémie tout en faisant face aux enjeux sociétaux suivants :
\begin{itemize}
    \item Les règles de confinement sont jugées trop strictes et préjudiciables à la santé mentale. Il est essentiel de trouver des mesures qui ne soient pas trop contraignantes pour la population mais toujours efficaces contre l'épidémie.
    \item La violation des règles par quelques citoyens pourrait mettre en péril la stratégie de contrôle de l'épidémie, menaçant ainsi l'ensemble de la population. Il est donc indispensable de faire respecter efficacement les restrictions sanitaires, sans surcharger les policiers de missions supplémentaires. En outre, des règles contraignantes pourraient ne pas toujours être bien acceptées et des efforts seront nécessaires pour empêcher la détérioration des relations entre la police et les citoyens.
    \item Les personnes asymptomatiques pourraient propager l'épidémie sans le savoir, ce qui rendrait insuffisant le dépistage et la mise en quarantaine des seules personnes symptomatiques. Cependant, il sera difficile d'imposer des restrictions aux personnes en bonne santé qui ne se sentent pas concernées. Les restrictions sanitaires doivent être adaptées à l'état de santé des citoyens.
\end{itemize}
La description du contexte ne mentionne volontairement jamais le nom du virus responsable de l'épidémie à Wafer city, et évite toute allusion au COVID-19. En effet, le jeu vise à discuter des implications sociétales de l'IA dans un contexte plus large, et devrait rester d'actualité même après la fin de la pandémie actuelle. Par ailleurs, nous voulions également prendre de la distance par rapport à un contexte très sensible, car l'épidémie de COVID-19 n'est pas encore terminée.

\subsection{Sowana}
La société Sowana est spécialisée dans l'Intelligence Artificielle et a déjà travaillé avec la ville sur d'autres sujets, comme les transports en commun ou la santé. Sowana a répondu à l'appel à projets de Wafer city concernant l'épidémie avec trois solutions : \textit{Eye'Wana}, \textit{Wana'Like} et \textit{Wana'Pass}, qui seront détaillées dans la prochaine section. Ces trois solutions font appel à diverses technologies d'Intelligence Artificielle et permettent donc d'éduquer les participants sur leur fonctionnement technique et sur leurs risques.
\begin{itemize}
    \item \textbf{Eye'Wana} utilise la vidéo-surveillance et des algorithmes de reconnaissance faciale pour garantir l'égalité devant la loi : tout le monde est contrôlé et verbalisé de la même manière, sans biais humains.
    \item \textbf{Wana'Like} utilise des algorithmes de recommandation qui analysent le profil des utilisateurs et les données d'affluence, pour recommander des sorties plus sécurisées dans des lieux peu fréquentés.
    \item \textbf{Wana'Pass} est un passeport sanitaire à points, qui analyse les données médicales pour attribuer à chacun des droits d'accès aux lieux à risque adaptés à son état de santé actuel.
\end{itemize}

\subsection{Rôles représentés au conseil municipal de Wafer city} \label{ssec:roles}

Les différentes solutions proposées par Sowana ont toutes des avantages et des inconvénients, et seront plus ou moins acceptables pour différents groupes d'utilisateurs. Dans notre jeu-débat, les participants ne jouent pas en leur propre nom, mais assument un rôle en tant que membre d'un groupe spécifique de citoyens. L'idée de faire jouer aux utilisateurs un rôle différent n'est pas nouvelle et s'est avérée favoriser le changement de perspective \cite{jarvis2002role,resnick1998diving}. Nous voulions aussi que les participants prennent de la distance avec leur personnage pour garantir un débat plus apaisé, d'autant plus sur un sujet très sensible alors que l'épidémie de COVID-19 n'est toujours pas terminée.

\begin{figure}[hbt]
    \centering
    \includegraphics[scale=0.2]{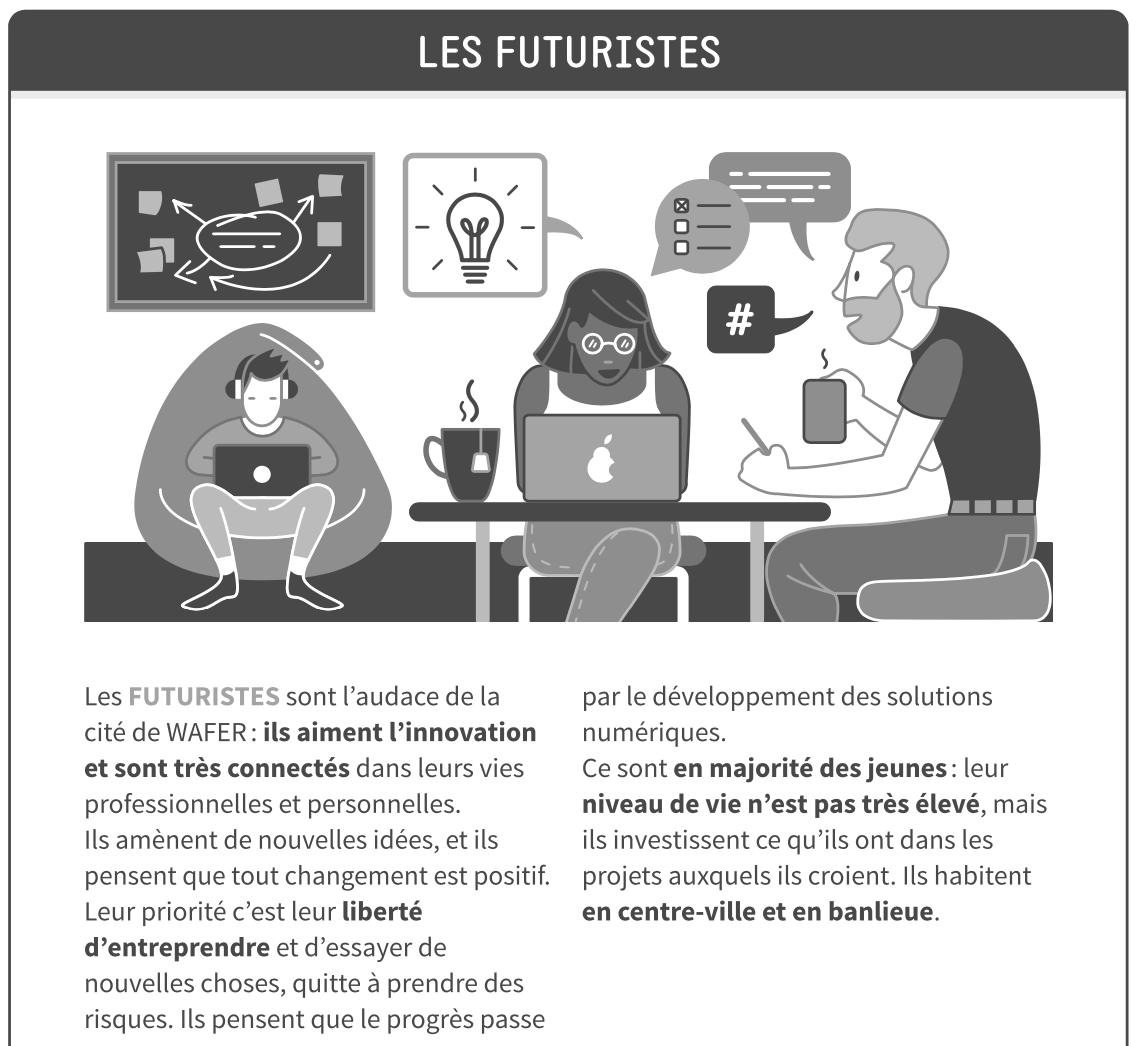}
    \includegraphics[scale=0.2]{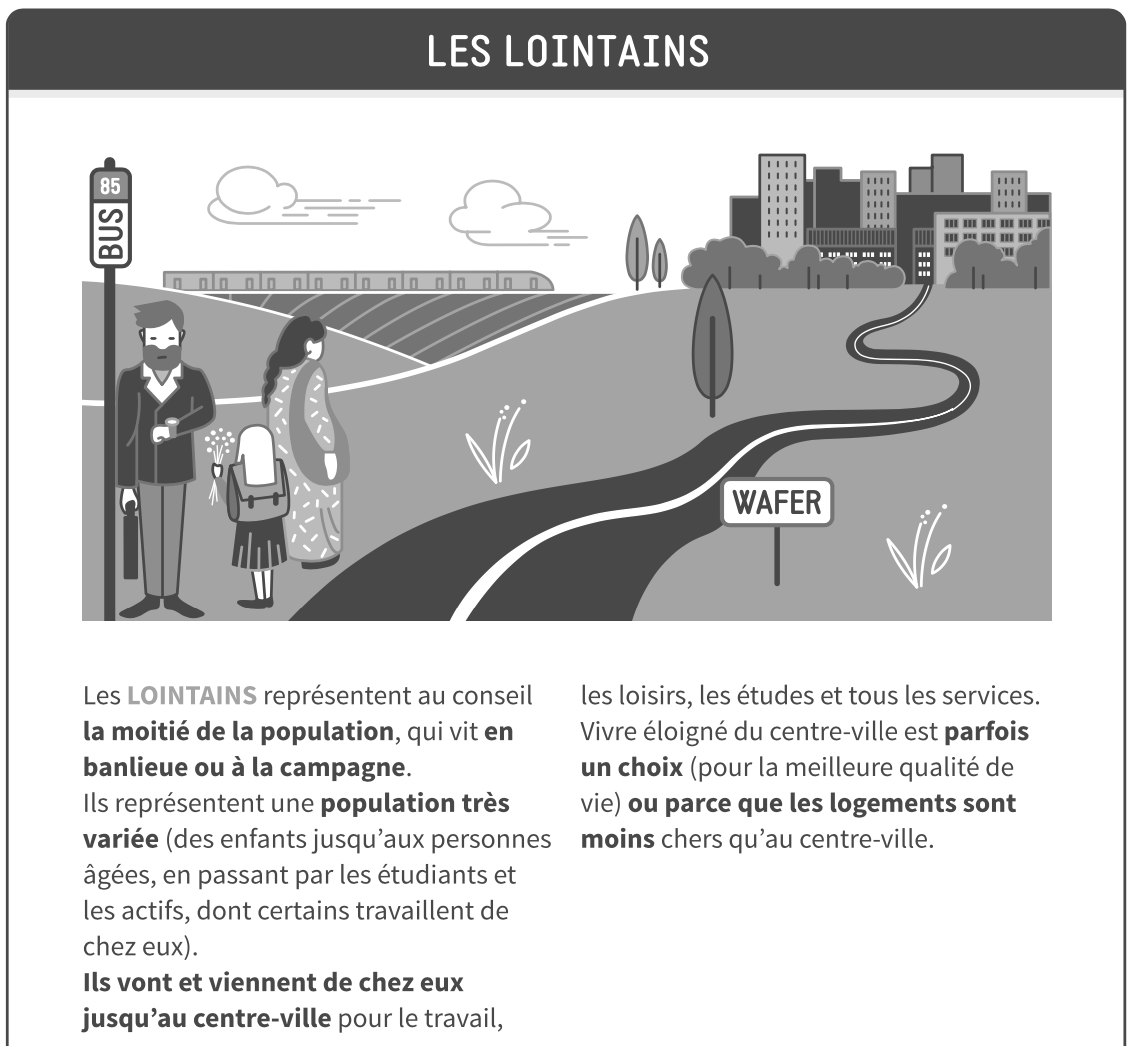}
    \caption{Extrait des cartes du jeu décrivant 2 des rôles. Illustration \copyright~ Marie Jamon}
    \label{fig:roles}
\end{figure}

\noindent Les participants à notre jeu sont donc répartis aléatoirement en 5 petits groupes de citoyens :
\begin{itemize}
    \item Les \textbf{Central Wafer} : les travailleurs, commerçants, artisans du centre-ville, ils tiennent à leur liberté, à leur mobilité et celle de leurs clients. Ils ont des revenus corrects mais peu de temps libre. 
    \item Les \textbf{Alter Wafer} : plutôt jeunes à sensibilité écologiste ou alter-mondialiste, focalisés sur le bien-être de tous et la protection de la nature. Ils vont au travail à pieds ou vélo, et demandent des garanties avant d'adopter de nouvelles technologies.
    \item Les \textbf{Seniors} : des personnes âgées, de plus de 60 ans, retraitées, focalisées sur le maintien de leur autonomie. Ils peuvent vivre en ville ou à la campagne.
    \item Les \textbf{Futuristes} : surtout des jeunes, amateurs de nouvelles technologies, très connectés et actifs en ligne, ils adorent l'innovation mais ont des revenus limités. 
    \item Les \textbf{Lointains} : groupe très hétérogène, vivant en dehors du centre-ville (banlieue ou zone rurale),~pour une meilleure qualité de vie ou pour les loyers plus bas, ils font donc de nombreux trajets vers la ville.
\end{itemize}

Les membres de chaque groupe débattront entre eux pour allouer un total de 5 points entre les 3 solutions, en gardant à l'esprit le profil qu'ils représentent. Chaque groupe-rôle convergera ainsi vers des scores (très) différents pour les solutions, scores qui seront additionnés pour déterminer la solution gagnante. Ces scores devront être argumentés, et donneront aussi lieu à une discussion lors du débriefing.

\subsection{Matériel de jeu} \label{ssec:matos}

Le matériel de jeu\footnote{Le matériel de jeu et le livret sont disponibles gratuitement en ligne \anonymisation{\url{https://jeudebat.com/jeux/lintelligence-artificielle/}}{LIEN ANONYMISE}} comprend les éléments suivants :
\begin{itemize}
    \item Cartes décrivant les rôles, une carte différente à distribuer à chaque groupe ;
    \item Une fiche expliquant l'ordre du jour du conseil municipal, et les problèmes à résoudre ;
    \item Une fiche par solution, qui fournit l'argumentaire marketing de Sowana, des détails techniques sur l'IA sous-jacente, et le témoignage d'un citoyen destiné à soulever quelques premières interrogations ;
    \item Un guide du modérateur destiné aux professeurs qui superviseront le jeu. Ce guide contient des informations plus techniques pour approfondir les notions en cours, des définitions utiles, des liens vers de la documentation supplémentaire, des questions pour alimenter le débat, et des informations sur des solutions d'IA similaires déjà réelles à présenter lors du débriefing.
\end{itemize}
Ces éléments sont fournis lors des séances que nous organisons, mais ils peuvent également être imprimés par tout enseignant désireux de jouer le jeu avec ses élèves, de manière autonome.

\section{Les solutions d'Intelligence Artificielle} \label{sec:solutions}

Cette section détaille les trois solutions d'Intelligence Artificielle proposées par Sowana pour combattre l'épidémie à Wafer City. Nous indiquons le principe de la solution, les notions d'IA impliquées qu'elle permet d'introduire aux participants, ainsi que les questions éthiques qu'elle soulève. Ces questions éthiques seront éventuellement suggérées pendant la phase de réflexion, puis seront abordées en détails lors du débriefing.

\subsection{Eye'Wana}

\textit{Eye'Wana} s'appuie sur la vidéosurveillance pour s'assurer que les règles sanitaires sont respectées de la même manière par tous les citoyens à tout moment. Eye'Wana peut détecter diverses infractions aux règles sanitaires (non-port du masque ou non-respect de la distanciation physique ; violation du couvre-feu ou du confinement ; etc.), en identifier l'auteur, et lui notifier l'amende qu'il devra payer en conséquence.

\paragraph{Technologie.}
Cette solution repose sur trois briques : (i) un réseau de vidéosurveillance et une flotte de drones collectent les vidéos et les images des lieux publics, (ii) des algorithmes de vision par ordinateur extraient les visages de ces images, et (iii) des algorithmes de reconnaissance faciale identifient les individus à partir de leur visage.

\paragraph{Dans la réalité.} 
Les technologies de reconnaissance faciale sont de plus en plus utilisées (par la police dans certains pays, pour les contrôles d'identité dans certains aéroports, pour déverrouiller un téléphone, etc). Des technologies similaires à Eye'Wana ont été utilisées en Chine~\cite{Chun2020}, où un vaste réseau de caméras de vidéosurveillance avec drones et caméras thermiques, complété par la reconnaissance faciale, a été déployé pour mettre en œuvre un contrôle social pendant la pandémie de COVID-19. Mais des drones ont aussi été expérimentés en France pour surveiller les plages de Nice pendant le confinement\furl{https://news.trust.org/item/20200320105803-ztaq0}.

\paragraph{Bénéfice-risque.}
Eye'Wana promet de décharger les forces de police tout en évitant les préjugés et biais humains : tout le monde a la même probabilité d'être contrôlé et verbalisé en cas de non-respect des règles. Cette solution répond donc au besoin d'égalité devant la loi, mais au prix d'une forte intrusion dans la vie privée des personnes, puisqu'elle traque leurs moindres faits et gestes. Par ailleurs, la reconnaissance faciale n'est pas infaillible, et implique en fait un fort risque de discrimination \cite{bacchini2019race,najibi2020racial}, qui a d'ailleurs conduit à son interdiction récente dans certains états aux États-Unis. 
Elle pose aussi un certain nombre de questions éthiques \cite{martinez2019important}. 


\subsection{Wana'Like}

\textit{Wana'Like} est une application gratuite pour \emph{smartphone} destinée à aider les citoyens à éviter les zones surpeuplées où le risque de contamination est plus élevé. Sowana tient à jour une liste de lieux recommandés, vérifie s'ils respectent le protocole sanitaire, et peut les bannir si des contaminations sont signalées dans leur locaux. L'application analyse les goûts et intérêts de l'utilisateur à partir de son activité en ligne (publications sur les réseaux sociaux, requêtes sur le moteur de recherche Sowana...) pour créer un profil utilisateur. Elle utilise également des données de localisation en temps réel pour déduire la fréquentation actuelle des lieux. En combinant ces données, Wana'Like peut faire des suggestions de loisirs personnalisées mais sûres, et propose des réductions dans les lieux partenaires pour inciter davantage les utilisateurs à suivre les recommandations.

\begin{figure}[hbt]
    \centering
    \includegraphics[scale=0.15]{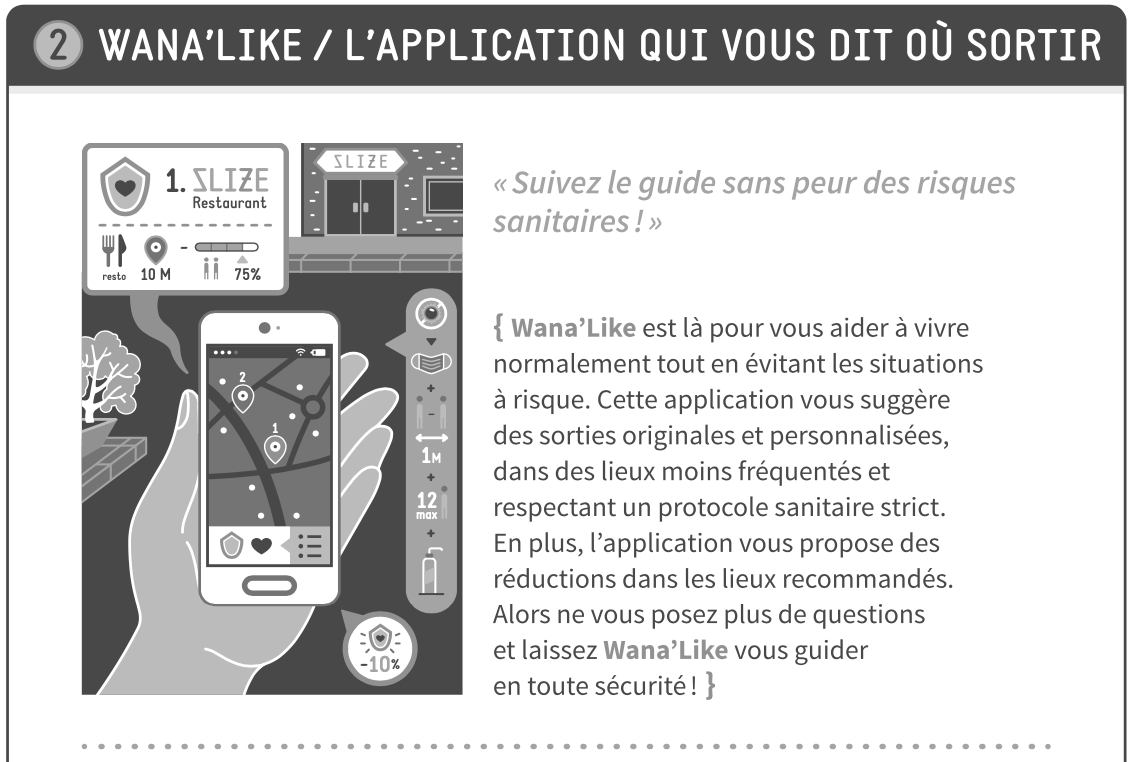}
    \includegraphics[scale=0.25]{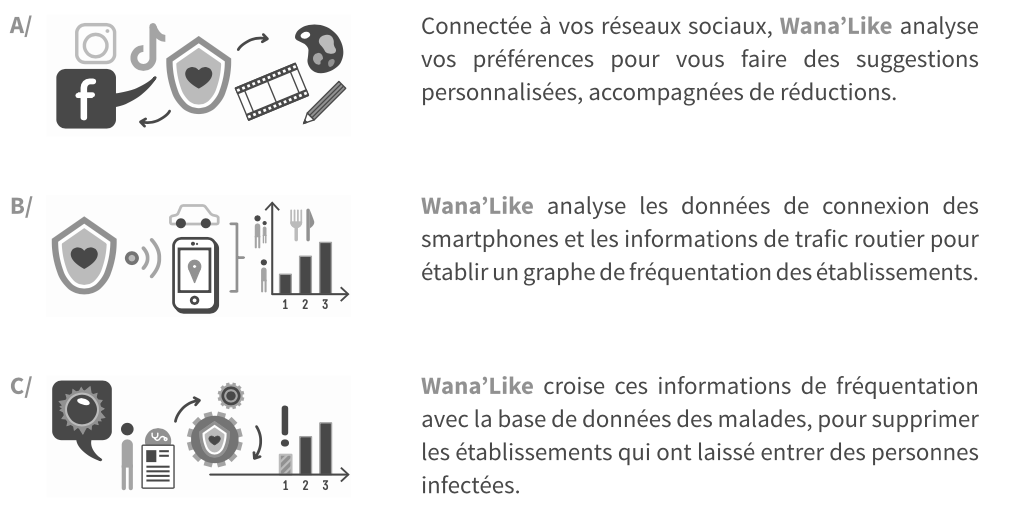}
    \caption{Extrait de la carte de jeu décrivant Wana'Like - illustration \copyright~ Marie Jamon}
    \label{fig:wanalike}
\end{figure}

\paragraph{Technologie.}
Cette solution met en lumière le fonctionnement d'algorithmes très utilisés dans la vie courante, notamment sur les réseaux sociaux, pour la recommandation de contenu, de vidéos à regarder, de films à aller voir au cinéma, d'objets à acheter, d'un restaurant, etc. 

Un algorithme de recommandation permet de filtrer des informations, des objets, afin de trouver automatiquement ce qui pourrait intéresser son utilisateur. Il fonctionne en 3 étapes: (i) recueillir de l'information sur l'utilisateur, soit de manière active (explicite) en lui demandant de classer, noter des éléments, soit de manière passive (implicite) en analysant son comportement en ligne, ses choix antérieurs ; (ii) construire et maintenir un profil de l'utilisateur contenant ses préférences ; (iii) proposer des recommandations, basées sur ses préférences pour les propriétés des éléments disponibles, ou sur les choix d'autres utilisateurs similaires, ou sur une combinaison de ces techniques.


\paragraph{Dans la réalité :} Wana'Like partage de nombreuses similitudes avec les réseaux sociaux, qui utilisent des algorithmes de recommandation pour faire des suggestions personnalisées à l'utilisateur, et l'orienter vers des contenus (posts, vidéos, photos) susceptibles de l'intéresser. Pendant la pandémie, on a aussi vu Google Maps proposer une mesure de l'affluence dans les lieux publics basée sur les données des smartphones pour recommander d'éviter certains lieux à certaines heures, ou encore le projet NoBis se basant sur le scan de QR codes en entrée et sortie d'un lieu référencé \cite{avanzi2021nobis}. 

\paragraph{Bénéfice-risque.}

Cette solution est beaucoup moins restrictive, mais potentiellement aussi moins efficace, et pose la question de l'accessibilité aux personnes ne possédant pas de \emph{smartphone}. En outre, malgré leur apparence inoffensive (les données sont fournies par l'utilisateur de son plein gré, dans le but d'en tirer un bénéfice), les algorithmes de recommandation posent question. En effet, ils sont très gourmands en données personnelles qui pourraient ensuite être utilisées à mauvais escient, comme pour manipuler les opinions lors d'élections \cite{aral2019protecting}. Ils ont aussi tendance à orienter vers des contenus de désinformation \cite{faucon2021recommendation}, comme les théories du complot ou anti-vaccination pendant la pandémie de COVID. Des travaux ont ainsi montré une mortalité accrue chez les personnes s'informant principalement sur les réseaux sociaux \cite{nieves2021infodemic}. Ces données personnelles pourraient aussi permettre de déduire des informations sensibles, comme l'orientation sexuelle à partir des photos \cite{wang2018deep} ou des liens Facebook \cite{jernigan2009gaydar}, ou encore la fréquentation de lieux sensibles d'après les données de localisation, comme les cliniques d'avortement (Google annonce d'ailleurs récemment\footnote{Voir par exemple : \url{https://www.lesnumeriques.com/vie-du-net/google-supprimera-les-donnees-des-personnes-visitant-des-cliniques-pratiquant-l-avortement-n187043.html}} supprimer automatiquement les données en cas de visites de tels lieux sensibles). 

Il existe donc de grands risques liés à l'utilisation commerciale des données personnelles (publicité), à la perte de vie privée, ou à la manipulation d'opinion. Il y a aussi un conflit d'intérêts entre l'utilisateur et la plateforme, puisque les algorithmes visent à maximiser le temps passé en ligne, au détriment de la santé (mentale) de l'utilisateur \cite{wells2021facebook}. Enfin, il y a un risque de discrimination des personnes ne possédant pas de smartphone.

\subsection{Wana'Pass}

\textit{Wana'Pass} est un passeport sanitaire à points. Il analyse quotidiennement le dossier médical de l'utilisateur et les données de ses capteurs (fréquence cardiaque, température...) pour en déduire son état de santé et son niveau de risque. Il lui attribue ensuite un certain nombre de points, les utilisateurs en meilleure santé recevant plus de points et les personnes infectées n'en recevant aucun. Ces points seront nécessaires pour accéder aux lieux publics (magasins, restaurants, musées, etc.), les endroits les plus risqués nécessitant plus de points d'accès. Chaque lieu devra vérifier et débiter les points d'accès nécessaires ; une fois qu'il n'a plus aucun point (ou tant qu'il est contagieux), l'utilisateur ne peut plus entrer dans ces lieux.

\paragraph{Technologie.}
Wana'pass peut être considéré comme le résultat des progrès de l'IA médicale~\cite{Petropoulos2020}, et montre une utilisation possible de l'IA pour analyser des données médicales. Les premiers algorithmes d'IA médicale utilisaient des systèmes experts pour coder les connaissances médicales et les réutiliser sur de nouveaux scénarios, pour aider par exemple à poser un diagnostic ou proposer un traitement. Les algorithmes d'apprentissage automatique permettent maintenant d'analyser d'énormes volumes de données, qu'un humain ne pourrait traiter, pour apprendre à résoudre une tâche sans être explicitement programmés pour.

\paragraph{Dans la réalité.} 
De nombreuses applications d'IA médicale existent déjà, notamment en radiologie, comme la détection de tumeurs sur des mammographies \cite{sheth2020artificial}. 
Pendant la pandémie de COVID, la France et d'autres pays ont mis en place un passeport sanitaire, mais basé uniquement sur le statut vaccinal, la guérison d'une infection au COVID, ou le résultat d'un test de dépistage. Ce passeport sanitaire manipule donc déjà des données médicales, privées. La mise en place récente du Dossier Médical Partagé en France rend une solution similaire encore plus crédible.
Dans un autre contexte, Wana'Pass rappelle aussi le système de crédit social chinois~\cite{Shen2019,liu2019multiple}.

\paragraph{Bénéfice-risque.}

Cette solution vise à répondre au problème d'un traitement équitable plutôt qu'égalitaire~: elle protège les personnes à risque et isole les personnes contagieuses, tout en laissant les personnes en bonne santé vivre plus normalement afin de protéger leur santé mentale. Cependant, elle pose la question de discriminer et d'isoler encore plus les personnes âgées sous prétexte de les protéger. Sa mise en place exige aussi des garanties fortes sur la protection des données médicales utilisées.
Par ailleurs, l'obligation de présenter un passeport vaccinal pour accéder à un certain nombre de lieux publics, pose le problème des discriminations créées par l'accès inégal aux vaccins \cite{tanner2021vaccine, gostin2021digital, kofler2020ten}.


\section{Débriefing} \label{sec:debrief}

L'objectif de ce jeu est d'éduquer les participants au sujet de l'Intelligence Artificielle, non seulement d'un point de vue technologique (le fonctionnement général des algorithmes sous-jacents) mais aussi d'un point de vue éthique (les questions soulevées par son utilisation). Le simple fait de jouer et débattre n'est cependant pas suffisant, et il est essentiel de conduire un bon débriefing pour permettre aux joueurs d'assimiler le message. 

Le débriefing est en effet une part importante d'un jeu sérieux, et indispensable à l'apprentissage \cite{crookall2010serious, whalen2018all, dieleman2006games}. Il poursuit plusieurs buts : clarifier les concepts manipulés pendant la session de jeu, afin d'améliorer l'apprentissage~; relier ces concepts, simplifiés pendant le jeu, avec la complexité du monde réel, pour s'assurer que les participants puissent appliquer leurs connaissances en situation réelle~; et partager les expériences et la réflexion entre les participants. Le débriefing doit être conduit par un modérateur informé, avec à l'esprit un objectif pédagogique clairement défini. Les paragraphes qui suivent décrivent certaines phases du débriefing de notre jeu-débat.

\subsubsection{Partage et débat sur les scores}

À la fin du débat, chaque groupe donne les notes qu'il a attribuées à chaque solution. L'animateur reporte au tableau les scores de tous les groupes, ainsi que leurs principaux arguments pour justifier ces scores (voir un exemple en Tableau~\ref{tab:board}). La solution avec le score total le plus élevé "gagne" ce premier tour de scrutin.



\strut \qquad \qquad 
\begin{table}[ht]
    \flushright
    \begin{tabular}{|c|c|c|c|}
    \hline
       Solution & S1 Eye'Wana & S2 Wana'Like & S3 Wana'Pass \\
       \hline
        Alter-glob & 0 & 3 (+) discounts & 2 (-) unfair  \\
                   &   & (+) normal life & \\
                         & (-) privacy & (+) control over data & (-) mental health\\
        \hline
        Shops & 4 (+) stay open & 0 (-) lose clients & 1 (-) lose clients \\
        \hline
        Remote & 1 & 3 (-) privacy & 1\\
                        &   & (+) discounts & (-) unfair \\
        \hline
        Techno & 2 (+) fair, jobs & 2 (+) freedom & 1 \\
               &                  & (+) discounts & \\
        \hline
        Elderly & 3 (+) health & 2 (-) no smartphone & 0 (-) autonomy \\
            & (+) live on &    & (-) privacy \\
        \hline
        \hline
        TOTAL & 10 & 10 & 5 \\
        \hline
    \end{tabular}
    \caption{Transcription du tableau de jeu en Fig~\ref{fig:board}}
    \label{tab:board}
\end{table}

\strut \\

L'animateur déclenche alors un débat entre partisans et adversaires des différentes solutions. Par exemple, les membres d'un groupe ayant rejeté une certaine solution sont invités à essayer de convaincre les membres d'un autre groupe qui eux la soutiennent, et vice versa. S'ils réussissent, les scores au tableau peuvent ensuite être mis à jour, et la solution choisie peut même changer à la suite de ce débat (la section~\ref{sec:rex} donne un exemple d'un tel débat).

Comme recommandé par \cite{dieleman2006games}, le débat est encadré par un modérateur, qui assure un débat juste, serein et équilibré entre les élèves. Jouer un rôle qui ne correspond pas nécessairement à ses propres opinions aide également les élèves à prendre de la distance avec l'argumentation et à s'exprimer plus librement.

\subsection{Échelles d'évaluation}

Afin d'approfondir la comparaison entre les différentes solutions technologiques, le modérateur pose ensuite quelques questions précises. Il guide les participants pour évaluer les solutions d'IA sur les échelles suivantes, importantes à prendre en compte (inspirées de ~\cite{Castelluccia2020})~:
\begin{itemize}
    \item Leur \textbf{efficacité} (ici contre l'épidémie)~: l'équilibre bénéfice-risque est un indicateur clé, employé par les agences de protection de la vie privée avant d'autoriser une technologie utilisant des données personnelles, et par les institutions de santé avant d'autoriser un médicament ;
    \item Leur impact sur les \textbf{libertés individuelles}~: à quel point sont-elles contraignantes ou restrictives ? Combien de liberté est sacrifiée ? Est-ce que cela vaut le coup? Ont-elles des effets secondaires négatifs, par exemple sur la santé mentale ?
    \item Leur impact sur la \textbf{vie privée}~: quelles sortes de données personnelles collectent-elles ? Ces données sont-elles sensibles ? Qui y a accès ? Que se passe-t-il en cas de fuite ?
    \item Leur \textbf{coût} économique et écologique : combien cela coûte ? Qui va payer pour le déploiement de ces technologies ? Quelle pollution est générée pour les créer, maintenir, utiliser, recycler ? 
    \item Leur \textbf{accessibilité} pour tous, et les risques de discrimination. Assurent-elles une stricte égalité (tout le monde est traité de la même façon) ? Sont-elles équitables (les règles sont adaptées à chacun) ?
    \item Les \textbf{risques}  d'erreurs et de mauvaise utilisation : quel risque d'échec de ces technologies ? Quel impact en cas d'erreur ? Quel risque de mauvaise utilisation ? Peuvent-elles être détournées pour servir un autre but que celui initialement prévu ? Si oui, avec quelles conséquences ?
\end{itemize}

\subsection{Contextualisation}


Pour faire le lien avec la réalité, les modérateurs illustrent enfin comment les différentes solutions virtuelles évoquées pendant le jeu renvoient en fait à des solutions bien réelles déjà mises en place dans certains pays (cf Section~\ref{sec:solutions}). Ces exemples illustrent très concrètement les dangers potentiels d'une dérive de l'usage utile à l'usage forcé de l'IA, et aident les participants à comprendre notre message qu'ils sont tous concernés par l'IA et son utilisation. 

\section{Retour d'expérience} \label{sec:rex}

Ce jeu a été joué pour la première fois avec 6 groupes de 3 classes de lycée lors de la Fête de la Science (pour un total de 103 élèves) en Octobre 2021, comme un premier test grandeur nature. Il a ensuite été testé auprès d'environ 80 étudiants d'une école d'ingénieurs. Pour ces sessions nous avons imprimé le matériel de jeu mentionné plus haut (paragraphe~\ref{ssec:matos}), et assuré nous-mêmes l'animation et la modération. 
Dans les paragraphes suivants, nous présentons quelques retours d'expérience issus de ces premières sessions test.

\subsection{Mise en place}
Pendant la Fête de la Science, chaque session a été jouée avec un groupe d'environ 15-17 lycéens (une demi-classe), répartis dans les 5 groupes de rôles, plus un de leurs professeurs présents soit en tant qu'observateur, soit en tant que participant. La séance d'école d'ingénieur s'est déroulée avec les 80 élèves réunis, répartis en 10 groupes (2 par rôle). 
Chaque séance durait 1h30 à 2h et était encadrée par 2 modérateurs qui animaient les échanges en restant neutres, c'est-à-dire sans les orienter ni faire de suggestions. 


\subsection{Un exemple de débat} \label{ssec:exdeb}
Nous détaillons ici le débat qui s'est tenu avec l'un des groupes. Nous trouvons cet exemple particulièrement intéressant car il a mené à plusieurs changements de la solution sélectionnée par le conseil municipal. 

Après le premier tour de débat (Tableau~\ref{tab1}) Wana'Pass gagne avec 10 points, suivie de près par Eye'Wana avec 8 points, puis Wana'Like avec 7 points. 
\begin{table}[!ht]
    \centering
    \footnotesize
    \strut \hspace*{-10pt}
    \begin{tabular}{|r|c:c|c:c|c:c|}
    \hline 
       & \multicolumn{2}{c|}{\textbf{Eye'Wana}} & \multicolumn{2}{c|}{\textbf{Wana'like}} & \multicolumn{2}{c|}{\textbf{Wana'Pass}} \\ \hline
  
   \textbf{Group} & S & Arguments & S & Arguments & S & Arguments \\ \hline
   
\multirow{2}{*}{Seniors} & \multirow{2}{*}{4}& (+) Pas besoin smartphone& \multirow{2}{*}{1}&  & \multirow{2}{*}{0} &  \\
                        &  &     &  & (-) Nécessite smartphone     &   &    (-) Pas de points si âgés  \\ \hline 

\multirow{2}{*}{Lointains} & \multirow{2}{*}{1}& (+) Force respect des règles& \multirow{2}{*}{0}&  & \multirow{2}{*}{4} & (+) Simple \\
                        &  & (-) Pas de caméra hors centre-ville  &  & (-) Risque de manipulation  &   &     \\ \hline 

\multirow{2}{*}{Centraux} & \multirow{2}{*}{3}& (+) Égalité& \multirow{2}{*}{1}& (+) Liberté & \multirow{2}{*}{1} &  \\
                        &  & (+) Bon pour les affaires &  & (-) Inégalité entre magasins &   &  (-) Perte de clients \\ \hline 

\multirow{2}{*}{Alter} & \multirow{2}{*}{0}& (-) Pollution& \multirow{2}{*}{2}&  & \multirow{2}{*}{3} & (+) Efficacité \\
                        &  & (-) Perte vie privée &  & (-) Besoin garanties sur données &   & (-) Perte vie privée    \\ \hline 

\multirow{2}{*}{Futuristes} & \multirow{2}{*}{0}& & \multirow{2}{*}{3}& (+) Gratuit & \multirow{2}{*}{2} & (+) Équitable \\
                        &  & (-) Perte de liberté  &  & (+) Coupons de réduction &   &     \\ \hline

\textbf{Total}        & 8 & &7  & &10 &  \\ \hline
    \end{tabular}
    \caption{Exemple de tableau après un tour de discussion~: scores (S), et arguments pour/contre chaque solution.} 
    \label{tab1}
\end{table}
Mais nous avons ensuite animé un débat entre les participants des groupes en désaccord sur les différentes solutions, débat qui a conduit a plusieurs revirements. Le Tableau~\ref{tab2} illustre l'évolution des scores totaux de chaque solution au fil des quatre tours du débat, avec le score de la solution gagnante en bleu : on voit ainsi qu'à l'issue du deuxième et troisième tour, c'est Eye'Wana qui l'emporte. Nous détaillons le contenu du débat ci-dessous.

\begin{table}[!ht]
    \centering
    \begin{tabular}{|c|c|c|c|}\hline
Round      & \textbf{Eye'Wana} & \textbf{Wana'like} & \textbf{Wana'Pass} \\ \hline
 \hline
1         & 8&7& \textbf{\underline{10}} \\ \hline
2         & \textbf{\underline{9}}&8&8\\ \hline  
3        & \textbf{\underline{11}}&5&9 \\ \hline
4        & 11&1&\textbf{\underline{13}} \\ \hline
    \end{tabular}
    \caption{Évolution des résultats du vote au fil de 4 tours de débat.}
    \label{tab2}
\end{table}


Le groupe des Seniors étant totalement opposé à Wana'Pass (qui ne leur accorderait probablement que très peu de points pour sortir, leur âge étant considéré comme un facteur de risque), il leur a été demandé d'essayer de convaincre d'autres groupes de ne pas choisir cette solution. Ils ont fait valoir que Wana'Pass était inéquitable, et ont convaincu les Centraux (commerçants) de déplacer le seul point qu'ils donnaient à Wana'Pass vers Eye'Wana (pour préserver leur clientèle âgée), et les Futuristes de déplacer 1 point de Wana'Pass vers Wana'Like. A l'issue de ce deuxième tour, Eye'Wana prend donc la tête avec 9 points (\emph{S4-L1-C4-A0-F0}), en légère avance sur les 2 autres solutions marquant 8 points chacune (Wana'Like \emph{S1-L0-C1-A2-F4}, Wana'Pass \emph{S0-L4-C0-A3-F1}).


Les Lointains ont ensuite été invités à apporter leurs arguments contre Wana'Like, solution à laquelle ils n'attribuaient aucun point. Ils ont fait valoir que cette application pouvait manipuler ses utilisateurs. En conséquence, les Seniors et les Centraux ont retiré 1 point à Wana'Like (leur seul point sur cette solution) et l'ont transféré à Eye'Wana (qui concentre maintenant leurs 5 points), tandis que les Futuristes ont transféré 1 point de Wana'Like à Wana'Pass (inversant ainsi leur changement précédent). A l'issue de ce troisième tour, Wana'Like n'a alors plus que 5 points (\emph{S0-L0-C0-A2-F3}), Wana'Pass remonte à 9 points (\emph{S0-L4-C0-A3-F2}), et Eye'Wana prend la tête avec 11 points (\textit{S5-L1-C5-A0-F0}).


Enfin, les Futuristes et les Alter, farouchement opposés à Eye'Wana (pour ses atteintes aux libertés, et parce que leurs revenus plus faibles les rendaient plus sensibles aux amendes, jugées inéquitables) ont réalisé qu'ils pouvaient changer le résultat : les Futuristes ont repris les 3 points attribués à leur solution favorite mais forcément perdante (Wana'Like) pour les reporter vers leur deuxième option Wana'Pass, afin d'empêcher Eye'Wana de gagner. Les Alter ont aussi reporté 1 point de Wana'Like (sur les 2 qu'ils lui attribuaient) vers Wana'Pass. On obtient donc un total inchangé de 11 points pour Eye'Wana (principalement grâce aux Seniors et Centraux, \textit{S5-L1-C5-A0-F0}), mais Wana'Like n'a plus qu'un unique point (des Alter : \textit{S0-L0-C0-A1-F0}), tandis que Wana'Pass rassemble 13 points (4 points des Lointains, 4 des Alter, 5 des Futuristes, et aucun de la part des Seniors et Centraux qui ont plébiscité Eye'Wana : \textit{S0-L4-C0-A4-F5}).


%
Il est intéressant de noter que parmi les différentes parties test que nous avons animées (voir les statistiques en section~\ref{sec:quanti}), c'est une rare occurrence de défaite totale pour Wana'Like (1 point sur un total possible de 25), et la seule occurrence d'une victoire de Wana'Pass.

\paragraph{Remarque.} On voit bien qu'un tel débat pourrait continuer encore, ce qui n'a pas été le cas faute de temps. Le but du jeu n'est effectivement pas de dégager une solution 'optimale' (il n'y en a précisément aucune) mais de faire ressentir aux participants le fragile équilibre bénéfice-risque des technologies, et comprendre les points de vue différents de personnes avec un profil différent. Nous voulons leur montrer que tout point de vue peut être valable, mais doit être justifié par des arguments pour l'expliquer et convaincre les autres. 

\subsection{Résultats quantitatifs} \label{sec:quanti}

Nous considérons ici les 6 sessions animées pendant la Fête de la Science en Octobre 2021, avec un total de 103 lycéens. Ces sessions sont en effet plus représentatives du public cible de notre jeu débat. Les résultats quantitatifs (résumés au Tableau~\ref{tab:wins}) sont les suivants~:
\begin{itemize}
    \item Wana'Like a été sélectionnée 3 fois, en particulier pour les avantages financiers (coupons de réduction) et pour son aspect moins contraignant par rapport aux autres solutions.
    \item Lors d'une autre session, Eye'Wana et Wana'Like l'emportent à égalité. Wana'Pass a été rejeté pour sa discrimination envers les plus âgés. 
    \item Eye'Wana est sélectionné seul lors d'une autre session, où il a été considéré plus simple à déployer (pas besoin de smartphone) et parce qu'il force l'égalité entre tous devant les règles. Dans toutes les autres sessions il a été jugé trop intrusif et farouchement rejeté.
    \item Wana'Pass n'a été sélectionné qu'une seule fois, après le débat résumé ci-dessus, mais plus pour faire barrage à Eye'Wana que par réel choix.
\end{itemize}


\begin{table}[hbt]
    \centering
    \begin{tabular}{|c|c|c|c|}
    \hline
         & Wana'Like & Wana'Pass & Eye'Wana \\
         \hline
         \hline
        \#Sélections & 4 & 1 & 2 \\
        \hline
    \end{tabular}
    \caption{Nombre de victoires par solution sur 6 sessions (total de 7 car une session a donné un match nul)}
    \label{tab:wins}
\end{table}



\subsection{Retours qualitatifs}

Nous avons reçu un retour globalement positif de la part de tous les étudiants et de leurs professeurs. Dans cette section, nous discutons de quelques commentaires spécifiques faits par les joueurs après les différentes sessions. Cela souligne l'importance de la phase de débriefing qui permet de tels échanges avec les participants, pendant lesquels ils prennent une certaine distance par rapport aux événements débattus lors du jeu.

\paragraph{Lien avec le COVID :} malgré nos efforts pour créer une ville et une épidémie "virtuelles", la plupart des participants ont immédiatement fait référence au "COVID" dans leurs discussions. L'un d'eux a même mentionné être opposé au vaccin COVID, ce qui a prouvé l'importance de garder une certaine distance par rapport au contexte actuel afin d'assurer un débat apaisé et ciblé. Cependant, de telles prises de position n'ont eu lieu que lors des débriefings et n'ont pas influencé les débats, puisque les joueurs devaient alors s'en tenir à leurs rôles. De plus, nous pensons que ce lien s'atténuera avec le temps, à mesure que l'épidémie actuelle s'estompera, tandis que notre jeu restera d'actualité.

\paragraph{Hypothèse d'infallibilité :} les étudiants supposaient généralement que les technologies proposées étaient infaillibles et, par conséquent, ils ne tenaient pas compte des effets néfastes d'une potentielle erreur. Cela confirme la nécessité d'enseigner les bases de l'IA ainsi que son risque de biais et d'erreurs. Il est nécessaire de savoir comment les algorithmes fonctionnent pour comprendre comment ils pourraient également échouer, et il est essentiel de comprendre l'impact de telles erreurs (discriminations, etc).

\paragraph{Sécurité vs liberté :} les étudiants se sont beaucoup plus focalisés sur la protection de leur liberté que sur la sécurité sanitaire et donc la lutte efficace contre le virus. Même si les débats ont évoqué l'efficacité des solutions proposées, celle-ci est passée après des arguments de perte de vie privée ou de liberté de mouvement, et en fait après tout autre argument. Cela n'a rien d'étonnant dans une tranche d'âge qui souvent ne se sent pas directement concernée par le risque sanitaire. Cela montre également un manque d'insistance sur la nécessité d'efficacité des solutions dans nos instructions de jeu. Nous avons depuis amélioré ce point en demandant explicitement aux participants de comparer les solutions sur les différentes échelles listées ci-dessus, y compris leur efficacité. Il s'agit ainsi de discuter s'ils préfèrent accepter une solution très contraignante (et efficace) pour une courte durée, ou une solution moins contraignante (mais moins efficace) qu'il faudra probablement maintenir plus longtemps.

\paragraph{Surveillance stricte vs diffuse :} bien que voulant protéger leur liberté et leur vie privée, la plupart des étudiants étaient soit inconscients, soit fatalistes, vis-à-vis de la surveillance déjà exercée sur eux par les algorithmes des réseaux sociaux. La plupart des étudiants ont considéré que les applications les plus intrusives étaient Wana'Pass du fait de l'utilisation de données médicales, ou Eye'Wana car il filme les gens en permanence. Ils sous-estiment complètement les risques liés aux algorithmes de recommandation, puisqu'ils divulguent déjà volontiers de grandes quantités d'informations personnelles sur les réseaux sociaux. Cela montre que l'éducation à l'IA devrait également insister sur la valeur des données personnelles et sur ce qu'il est possible d'en faire. De nombreuses références sont donc fournies dans le guide de l'animateur pour insister sur ce point.

\paragraph{Public cible :} ayant animé des sessions à la fois avec des lycéens (15 à 18 ans) et des élèves ingénieurs (22 à 24 ans), nous avons eu des retours assez différents. Les commentaires des participants ont tendance à suggérer que le jeu est mieux ciblé sur les élèves du secondaire, qui ont moins de connaissances techniques sur l'intelligence artificielle et sont heureux d'avoir l'occasion de discuter de la pandémie actuelle et de son impact sur leur vie. Au contraire, certains étudiants ingénieurs ont été déçus de l'accent mis sur les implications éthiques de l'IA alors qu'ils s'attendaient à acquérir plus de connaissances techniques. Étonnamment, certains ont signalé comme point négatif que les solutions semblaient trop irréalistes car "aucune n'était bonne" (alors même que nous leur avons montré des exemples réels de solutions similaires déjà en oeuvre dans d'autres pays). Incidemment, c'était précisément le message à retenir : aucune technologie d'IA n'est une solution parfaite, et tout le monde devrait être conscient des pièges avant de les accepter. Il semble donc important de cibler aussi ce public, même s'il est a priori moins ouvert, peut-être en changeant la forme d'intervention.

\paragraph{Didactique de l'informatique :} même si ce jeu-débat semble mettre l'accent sur les aspects sociétaux de l'informatique et de l'IA, il aborde aussi des points essentiels d'un point de vue technique. Tout d'abord, la phase d'introduction et le débriefing sont l'occasion de donner des définitions de différents concepts, souvent manipulés mais mal compris, comme l'IA ou les algorithmes. Ensuite, les solutions d'IA proposées sont toutes réalistes et basées sur des briques technologiques existantes, nous permettant d'introduire leur fonctionnement (reconnaissance faciale, fouille de données médicales, algorithmes de recommandation, etc). Enfin, la compréhension de ce fonctionnement permet de prendre conscience des risques associés : des erreurs sont possibles, des biais aussi, et les données collectées peuvent être mal utilisées. Le guide de l'animateur permet au professeur d'approfondir les notions souhaitées avec ses élèves. Nous espérons aussi avoir aiguisé leur curiosité et les pousser à s'informer plus sur le sujet. Des travaux sont en cours pour élaborer une séquence pédagogique à plus long-terme, associant un cycle de conférences sur l'IA, une session du jeu, des groupes de discussion, etc, avec un suivi par questionnaires de l'impact sur les élèves.


\section{Conclusion} \label{sec:cci}

Dans cet article, nous avons présenté un jeu sérieux sous forme de débat municipal visant à choisir des solutions d'IA pour lutter contre une épidémie, et les retours d'expérience des premières sessions de test. \anonymisation{Depuis avril 2022, ce jeu est maintenant disponible en libre accès sur le site de l'Arbre des Connaissances et a déjà été téléchargé 107 fois.}
%
Les participants sont répartis en 5 groupes de rôles (personnes âgées, commerçants, écologistes...) : changer ainsi de perspective, devoir comprendre d'autres points de vue et devoir argumenter pour défendre ses idées est un bon exercice pour eux. Mais les participants ont aussi appris sur l'intelligence artificielle et ses dangers, prenant conscience de sa faillibilité et de la valeur de leurs données personnelles. Ce jeu atteint donc son objectif de sensibilisation de la population, et la phase de débriefing s'est confirmée comme cruciale dans ce processus. Nous envisageons d'ailleurs d'étendre son utilisation en créant un cycle pédagogique plus long associant conférences et ateliers participatifs.

Par ailleurs, étant nous-mêmes \anonymisation{enseignants en Informatique, donnant des cours de cybersécurité ou d'Intelligence Artificielle}, le débriefing des débats avec les participants nous a également fourni des informations intéressantes sur leur façon d'appréhender l'IA, ses avantages, ses prétendues objectivité et infaillibilité. Cela a conforté notre idée que nous devons éduquer la prochaine génération sur ces technologies. Bien que l'IA puisse certainement aider à lutter contre une crise telle que la pandémie de COVID-19, elle n'est pas infaillible et, comme elle devient de plus en plus omniprésente dans nos vies, l'impact potentiel des erreurs devient également plus grave. Il est donc nécessaire que chacun connaisse les bases des algorithmes d'IA \cite{unesco} et puisse faire des choix éclairés quant à leur utilisation.



\section*{Remerciements}

\anonymisation{Les auteurs remercient l'association Arbre des Connaissances, en particulier Clara Fruchon et Camille Volovich, pour leur soutien tout au long du processus de conception du jeu et de sa diffusion\footnote{\anonymisation{Le jeu est disponible en téléchargement sur leur site à l'adresse \url{https ://jeudebat.com/jeux/lintelligence-artificielle/}}} ; ainsi que Marie Jamon pour l'illustration du matériel de jeu et du livret. Les auteurs remercient également l'équipe de médiation scientifique d'Inria Rhône-Alpes, en particulier Florence Polge-Cohen, pour son soutien financier et son aide à l'organisation des séances de jeu de la Fête de la Science. Enfin, nous remercions tous les élèves et enseignants qui ont participé aux séances de jeu.}{ANONYMISES}

\bibliographystyle{apalike}

\begin{thebibliography}{}

\bibitem[Adam and Arduin, 2022]{adam2022finding}
Adam, C. and Arduin, H. (2022).
\newblock Finding and explaining optimal screening strategies with limited
  tests during the covid-19 epidemics.
\newblock In {\em 19th International Conference on Information Systems for
  Crisis Response and Management ISCRAM}.

\bibitem[Adam et~al., 2022]{switch2022}
Adam, C., Jacquier, A., and Taillandier, F. (2022).
\newblock Un jeu sérieux pour sensibiliser aux enjeux d’une mobilité
  urbaine durable.
\newblock {\em Academic Journal of Civil Engineering (AJCE)}.

\bibitem[Adam and Lauradoux, 2022]{adam2022serious}
Adam, C. and Lauradoux, C. (2022).
\newblock A serious game for debating about the use of artificial intelligence
  during the covid-19 pandemic.
\newblock In {\em 19th International Conference on Information Systems for
  Crisis Response and Management (ISCRAM)}.

\bibitem[Aral and Eckles, 2019]{aral2019protecting}
Aral, S. and Eckles, D. (2019).
\newblock Protecting elections from social media manipulation.
\newblock {\em Science}, 365(6456):858--861.

\bibitem[Avanzi et~al., 2021]{avanzi2021nobis}
Avanzi, M., Coniglio, R., Cisotto, G., Giordani, M., and Ferro, N. (2021).
\newblock Nobis: A crowd monitoring service against covid-19.
\newblock In {\em IRCDL}, pages 126--137.

\bibitem[Bacchini and Lorusso, 2019]{bacchini2019race}
Bacchini, F. and Lorusso, L. (2019).
\newblock Race, again: how face recognition technology reinforces racial
  discrimination.
\newblock {\em Journal of information, communication and ethics in society}.

\bibitem[Castelluccia and M{\'{e}}tayer, 2020]{Castelluccia2020}
Castelluccia, C. and M{\'{e}}tayer, D.~L. (2020).
\newblock {Position Paper: Analyzing the Impacts of Facial Recognition}.
\newblock In {\em Privacy Technologies and Policy - 8th Annual Privacy Forum,
  {APF} 2020}, volume 12121 of {\em Lecture Notes in Computer Science}, pages
  43--57. Springer.

\bibitem[Chun, 2020]{Chun2020}
Chun, A. (2020).
\newblock {In a time of coronavirus, China’s investment in AI is paying off
  in a big way}.
\newblock {\em South China Morning Post}.
\newblock Last access 20 January 2022.

\bibitem[Cottineau and collective, 2020]{cottineau2020}
Cottineau, C. and collective, C. (2020).
\newblock Understanding the current covid-19 epidemic: one question, one model.
\newblock {\em RofASSS (Review of Artificial Societies and Social Simulation)}.

\bibitem[Crookall, 2010]{crookall2010serious}
Crookall, D. (2010).
\newblock Serious games, debriefing, and simulation/gaming as a discipline.
\newblock {\em Simulation \& gaming}, 41(6):898--920.

\bibitem[Dieleman and Huisingh, 2006]{dieleman2006games}
Dieleman, H. and Huisingh, D. (2006).
\newblock Games by which to learn and teach about sustainable development:
  exploring the relevance of games and experiential learning for
  sustainability.
\newblock {\em Journal of Cleaner Production}, 14(9-11):837--847.

\bibitem[Faucon et~al., 2021]{faucon2021recommendation}
Faucon, L., El-Mhamdi, E.-M., et~al. (2021).
\newblock Recommendation algorithms, a neglected opportunity for public health.
\newblock {\em Revue M{\'e}decine et Philosophie}, 4(2).

\bibitem[Gostin et~al., 2021]{gostin2021digital}
Gostin, L.~O., Cohen, I.~G., and Shaw, J. (2021).
\newblock Digital health passes in the age of covid-19: Are “vaccine
  passports” lawful and ethical?
\newblock {\em JAMA}, 325(19):1933--1934.

\bibitem[Graafland et~al., 2012]{graafland2012systematic}
Graafland, M., Schraagen, J.~M., and Schijven, M.~P. (2012).
\newblock Systematic review of serious games for medical education and surgical
  skills training.
\newblock {\em Journal of British Surgery}, 99(10):1322--1330.

\bibitem[Jackson and Back, 2011]{jackson2011teaching}
Jackson, V.~A. and Back, A.~L. (2011).
\newblock Teaching communication skills using role-play: an experience-based
  guide for educators.
\newblock {\em Journal of palliative medicine}, 14(6):775--780.

\bibitem[Jarvis et~al., 2002]{jarvis2002role}
Jarvis, L., Odell, K., and Troiano, M. (2002).
\newblock Role-playing as a teaching strategy.
\newblock {\em Strategies for application and presentation, staff development
  and presentation}.

\bibitem[Jernigan and Mistree, 2009]{jernigan2009gaydar}
Jernigan, C. and Mistree, B.~F. (2009).
\newblock Gaydar: Facebook friendships expose sexual orientation.
\newblock {\em First Monday}.

\bibitem[Kofler and Baylis, 2020]{kofler2020ten}
Kofler, N. and Baylis, F. (2020).
\newblock Ten reasons why immunity passports are a bad idea.
\newblock {\em Nature Publishing Group},
  https://www.nature.com/articles/d41586-020-01451-0\%C2\%A0.

\bibitem[Liu, 2019]{liu2019multiple}
Liu, C. (2019).
\newblock Multiple social credit systems in china.
\newblock {\em Economic Sociology: The European Electronic Newsletter},
  21(1):22--32.

\bibitem[Lu and Sun, 2022]{lu2022covid}
Lu, F. and Sun, Y. (2022).
\newblock Covid-19 vaccine hesitancy: The effects of combining direct and
  indirect online opinion cues on psychological reactance to health campaigns.
\newblock {\em Computers in human behavior}, 127:107057.

\bibitem[Martinez-Martin, 2019]{martinez2019important}
Martinez-Martin, N. (2019).
\newblock What are important ethical implications of using facial recognition
  technology in health care?
\newblock {\em AMA journal of ethics}, 21(2):E180.

\bibitem[Najibi, 2020]{najibi2020racial}
Najibi, A. (2020).
\newblock Racial discrimination in face recognition technology.
\newblock {\em Harvard Online: Science Policy and Social Justice}, 24.

\bibitem[Naudé, 2020]{naude2020}
Naudé, W. (2020).
\newblock {Artificial intelligence vs COVID-19: limitations, constraints and
  pitfalls}.
\newblock {\em AI \& Society}, pages 1--5.

\bibitem[Nieves-Cuervo et~al., 2021]{nieves2021infodemic}
Nieves-Cuervo, G.~M., Manrique-Hern{\'a}ndez, E.~F., Robledo-Colonia, A.~F.,
  and Grillo, A. E.~K. (2021).
\newblock Infodemic: fake news and covid-19 mortality trends in six latin
  american countries.
\newblock {\em Pan American Journal of Public Health}, 45:e44--e44.

\bibitem[Petropoulos, 2020]{Petropoulos2020}
Petropoulos, G. (2020).
\newblock {Artificial intelligence in the fight against COVID-19}.
\newblock Last accessed 20 January 2022.

\bibitem[Rebolledo-Mendez et~al., 2009]{rebolledo2009societal}
Rebolledo-Mendez, G., Avramides, K., De~Freitas, S., and Memarzia, K. (2009).
\newblock Societal impact of a serious game on raising public awareness: the
  case of floodsim.
\newblock In {\em Proceedings of the 2009 ACM SIGGRAPH symposium on video
  games}, pages 15--22.

\bibitem[Resnick and Wilensky, 1998]{resnick1998diving}
Resnick, M. and Wilensky, U. (1998).
\newblock Diving into complexity: Developing probabilistic decentralized
  thinking through role-playing activities.
\newblock {\em The Journal of the Learning Sciences}, 7(2):153--172.

\bibitem[Shen, 2019]{Shen2019}
Shen, C.~F. (2019).
\newblock {Social credit system in China}.
\newblock Technical report, City University of Hong Kong.

\bibitem[Sheth and Giger, 2020]{sheth2020artificial}
Sheth, D. and Giger, M.~L. (2020).
\newblock Artificial intelligence in the interpretation of breast cancer on
  mri.
\newblock {\em Journal of Magnetic Resonance Imaging}, 51(5):1310--1324.

\bibitem[Simonneaux, 2001]{simonneaux2001role}
Simonneaux, L. (2001).
\newblock Role-play or debate to promote students' argumentation and
  justification on an issue in animal transgenesis.
\newblock {\em International Journal of Science Education}, 23(9):903--927.

\bibitem[Taillandier and Adam, 2018]{taillandier2018games}
Taillandier, F. and Adam, C. (2018).
\newblock Games ready to use: A serious game for teaching natural risk
  management.
\newblock {\em Simulation \& Gaming}, 49(4):441--470.

\bibitem[Tanner and Flood, 2021]{tanner2021vaccine}
Tanner, R. and Flood, C.~M. (2021).
\newblock Vaccine passports done equitably.
\newblock {\em JAMA Health Forum}, 2(4):e210972--e210972.

\bibitem[{UNESCO}, 2021]{unesco}
{UNESCO} (2021).
\newblock Rapport de la commission sciences sociales et humaines (shs).
\newblock Technical Report 0000379920fre, {UNESCO}.

\bibitem[Wang and Kosinski, 2018]{wang2018deep}
Wang, Y. and Kosinski, M. (2018).
\newblock Deep neural networks are more accurate than humans at detecting
  sexual orientation from facial images.
\newblock {\em Journal of personality and social psychology}, 114(2):246.

\bibitem[Wells et~al., 2021]{wells2021facebook}
Wells, G., Horwitz, J., and Seetharaman, D. (2021).
\newblock Facebook knows instagram is toxic for teen girls, company documents
  show.
\newblock {\em The Wall Street Journal}.

\bibitem[Whalen et~al., 2018]{whalen2018all}
Whalen, K.~A., Berlin, C., Ekberg, J., Barletta, I., and Hammersberg, P.
  (2018).
\newblock ‘all they do is win’: Lessons learned from use of a serious game
  for circular economy education.
\newblock {\em Resources, Conservation and Recycling}, 135:335--345.

\bibitem[Wu and Lee, 2015]{wu2015climate}
Wu, J.~S. and Lee, J.~J. (2015).
\newblock Climate change games as tools for education and engagement.
\newblock {\em Nature Climate Change}, 5(5):413--418.

\end{thebibliography}

\end{document}